\documentclass[lettersize,journal]{IEEEtran}
\usepackage{amsmath,amsfonts}
\usepackage{algorithmic}
\usepackage{algorithm}
\usepackage{array}
\usepackage[caption=false,font=normalsize,labelfont=sf,textfont=sf]{subfig}
\usepackage{textcomp}
\usepackage{stfloats}
\usepackage{url}
\usepackage{verbatim}
\usepackage{graphicx}
\usepackage{booktabs}
\usepackage{orcidlink}
\usepackage{cite}
\begin{document}

\title{Toward a Machine Bertin: Why Visualization Needs Design Principles for Machine Cognition}

\author{Brian Keith-Norambuena~\orcidlink{0000-0001-5734-8962}%
\thanks{B. Keith-Norambuena is with the Department of Computing \& Systems Engineering, Universidad Cat\'{o}lica del Norte, Antofagasta, Chile.}%
}

\markboth{IEEE Transactions on Visualization and Computer Graphics}%
{Keith-Norambuena: When Visualization Meets Machine Cognition}

\maketitle

\begin{abstract}
Visualization's design knowledge---effectiveness rankings, encoding guidelines, color models, preattentive processing rules---derives from six decades of psychophysical studies of human vision.
Yet vision-language models (VLMs) increasingly consume chart images in automated analysis pipelines, and a growing body of benchmark evidence indicates that this human-centered knowledge base does not straightforwardly transfer to machine audiences.
Machines exhibit different encoding performance patterns, process images through patch-based tokenization rather than holistic perception, and fail on design patterns that pose no difficulty for humans---while occasionally succeeding where humans struggle.
Current approaches address this gap primarily by bypassing vision entirely, converting charts to data tables or structured text.
We argue that this response forecloses a more fundamental question: \emph{what visual representations would actually serve machine cognition well?}
This paper makes the case that the visualization field needs to investigate machine-oriented visual design as a distinct research problem.
We synthesize evidence from VLM benchmarks, visual reasoning research, and visualization literacy studies to show that the human-machine perceptual divergence is qualitative, not merely quantitative, and critically examine the prevailing bypassing approach.
We propose a conceptual distinction between \emph{human-oriented} and \emph{machine-oriented} visualization---not as an engineering architecture but as a recognition that different audiences may require fundamentally different design foundations---and outline a research agenda for developing the empirical foundations the field currently lacks: the beginnings of a ``machine Bertin'' to complement the human-centered knowledge the field already possesses.
\end{abstract}


\section{Introduction}
\IEEEPARstart{A}{rtificial} intelligence systems now routinely consume visualizations.
Automated reporting tools generate charts that feed into downstream analysis pipelines; multimodal AI assistants interpret dashboard screenshots; and agentic systems parse visual data summaries to inform decisions.
A growing body of research tests how well these systems understand the charts they encounter---and the results are revealing.

The evidence points in two directions simultaneously.
On one hand, Li et~al.~\cite{li2025visualization} find that providing scatterplots alongside raw data helps GPT-4.1 and Claude~3.5 analyze datasets more precisely, and visual reasoning research shows that prompting LLMs to generate visual representations improves spatial task performance~\cite{wu2024vot,hu2024sketchpad}.
On the other hand, VLMs perform poorly on charts designed for human comprehension: CharXiv~\cite{wang2024charxiv} reports a 33-point gap between human and GPT-4o performance, and multiple visualization literacy studies find that VLM error patterns are reliably \emph{distinct} from human error patterns---not just less accurate, but differently structured~\cite{bendeck2025empirical,verma2025chart6}.

These findings raise a question the visualization field has not yet systematically addressed: \emph{does machine cognition require a different design basis than human perception?}

The question matters because visualization's entire design knowledge---the accumulated output of six decades of research---is a science of human perception.
Bertin's visual variables~\cite{bertin1983semiology} characterize how graphical marks convey information to human perceivers.
Cleveland and McGill's effectiveness rankings~\cite{cleveland1984graphical} are derived from psychophysical experiments on human subjects.
Preattentive processing research~\cite{healey2012attention} identifies features that the human visual cortex detects without conscious attention.
Color models such as CIE~Lab are grounded in human cone cell responses.
Every design guideline the field has produced is, at its core, a claim about human perception.

Benchmark evidence increasingly suggests that this knowledge does not straightforwardly transfer to machine audiences.
VLMs exhibit different encoding performance patterns~\cite{mukherjee2025encqa}, and Poonam et~al.~\cite{poonam2025evaluating} replicate Cleveland and McGill's foundational experiments with Vision Transformers and find different effectiveness rankings.
Wang et~al.~\cite{wang2024dracogpt} show that LLM design preferences substantially diverge from human perceptual best practices.
The prevailing response has been to \emph{bypass vision}: convert charts to tables~\cite{liu2023deplot}, translate visual representations into structured text~\cite{masry2023unichart}, or use declarative grammars~\cite{satyanarayan2017vegalite}.

We argue that bypassing vision, while pragmatically useful, leaves a more fundamental question unasked: could visual representations be designed to serve machine cognition directly?
The fact that \emph{human-designed} charts fail for machines does not entail that \emph{all possible} visual representations fail for machines---and the emerging evidence that visual representations can benefit machine reasoning~\cite{li2025visualization,wu2024vot,hu2024sketchpad} suggests this distinction matters.

This paper argues that the visualization field needs to investigate machine-oriented visual design as a distinct research problem.
We propose a \textbf{conceptual distinction} between two forms of visualization:
\begin{enumerate}
  \item \textbf{Human-oriented visualization}: designed for human perception, following established design principles---the knowledge base the field already possesses.
  \item \textbf{Machine-oriented visualization}: designed for machine cognition, following design principles yet to be developed through systematic empirical investigation of how machines process visual information.
\end{enumerate}
This is not an engineering proposal for a specific pipeline architecture.
It is a recognition that if machines constitute a different kind of visual information consumer---which the evidence suggests they do---then the field's design knowledge needs a corresponding expansion.
What machine-oriented visualization ultimately looks like may be radically different from anything currently designed for humans; we do not claim to know the answer, only to identify the question.

This paper makes three contributions:
\begin{itemize}
  \item A theoretical analysis showing that visualization's knowledge base is grounded in human perceptual science (Section \ref{sec:knowledge_base}) and that benchmark evidence indicates it does not straightforwardly generalize to machine audiences (Section \ref{sec:machines_see_differently}).
  \item A critical examination of the prevailing approach of bypassing machine vision, arguing that it forecloses an important research direction (Section \ref{sec:against_bypassing}).
  \item An argument for why the field needs to investigate machine-oriented visual design, with a research agenda identifying the empirical questions that must be answered (Sections \ref{sec:machine_processing}, \ref{sec:two_layer}, and \ref{sec:agenda}).
\end{itemize}

\begin{figure*}[!t]
    \centering
    \includegraphics[width=\textwidth]{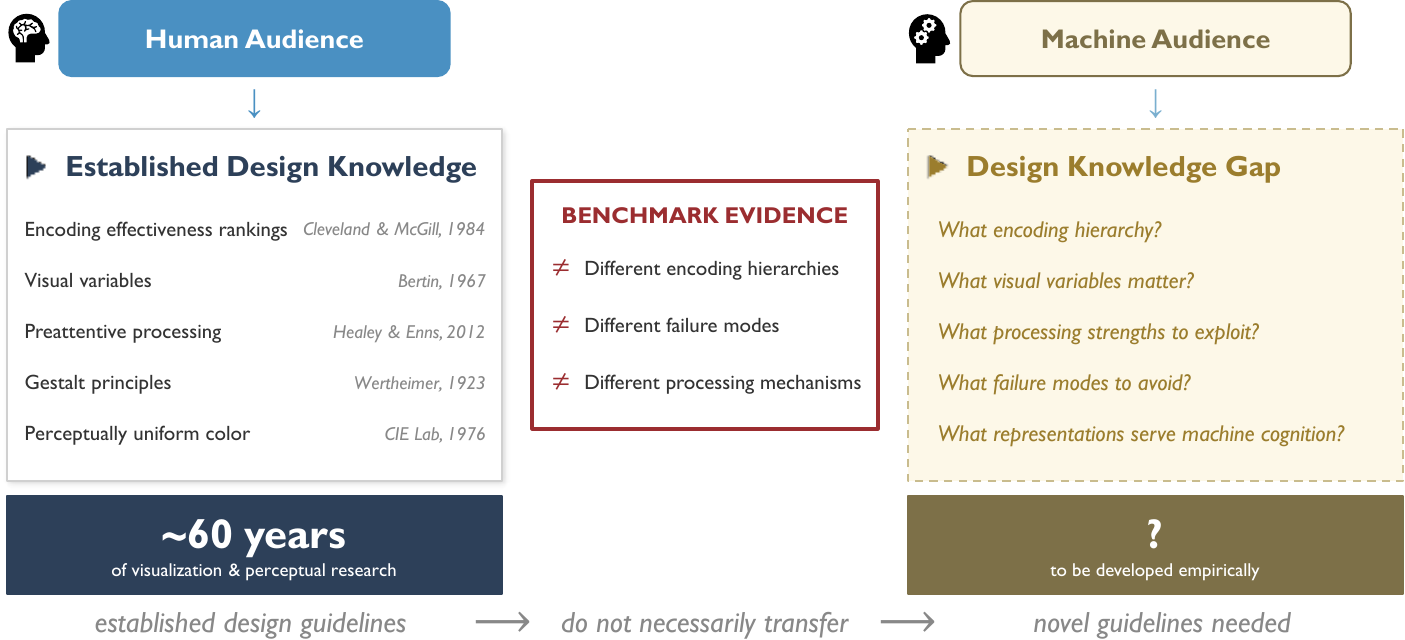}
    \caption{The conceptual distinction this paper draws. Visualization's design knowledge for human audiences (left) is grounded in six decades of perceptual science---a well-established body of research. Benchmark evidence (center) indicates that this knowledge does not necessarily transfer to machine audiences, whose visual processing differs in mechanism, encoding effectiveness, and failure modes. The corresponding design knowledge for machine-oriented visualization (right) does not yet exist. This paper argues that developing it is a research direction worth pursuing.}
    \label{fig:teaser}
\end{figure*}

\section{Related Work}
\label{sec:related_work}
Our analysis draws on and connects four bodies of research: VLM chart understanding benchmarks, human-machine perceptual comparison studies, structured and accessible visualization, and AI-visualization pipelines.

\subsection{VLM Chart Understanding}
The rapid development of vision-language models has spawned a large benchmark literature testing their ability to comprehend charts and visualizations.
ChartQA~\cite{masry2022chartqa} established the foundational benchmark for chart question answering, distinguishing between questions requiring visual extraction and those requiring logical reasoning.
Subsequent work has expanded in scope and diagnostic precision: CharXiv~\cite{wang2024charxiv} tests realistic chart comprehension from scientific papers; ChartMuseum~\cite{tang2025chartmuseum} introduces visual reasoning challenges; EncQA~\cite{mukherjee2025encqa} systematically isolates individual encoding channels; and ChartInsights~\cite{wu2024chartinsights} evaluates 19 models across low-level analytical tasks.

This literature has primarily adopted an \emph{evaluation} framing---measuring how well existing models perform on existing charts.
Our contribution reframes this evidence as a \emph{design} question: what do these benchmarks reveal about the information-processing properties of machine vision, and do those properties suggest the need for different visual design foundations?

\subsection{Human-Machine Perceptual Comparison}
A more recent strand of research directly compares human and machine visual processing of charts.
Bendeck and Stasko~\cite{bendeck2025empirical} evaluate GPT-4 on the VLAT visualization literacy assessment, finding it scores at the 16th percentile of human participants---but with error patterns that diverge qualitatively from human errors.
Verma et~al.~\cite{verma2025chart6} evaluate eight VLMs on six visualization literacy assessments and find that all models produce error patterns reliably distinct from human participants.
Poonam et~al.~\cite{poonam2025evaluating} replicate Cleveland and McGill's foundational graphical perception experiments with Vision Transformers and compare the resulting effectiveness rankings to those of human observers, finding differences.
Wang et~al.~\cite{wang2024dracogpt} extract visualization design preferences from LLMs and find they substantially diverge from human perceptual best practices.

This body of work provides the most direct evidence that the human-machine divergence in chart processing is qualitative, not merely a matter of lower accuracy.
The visualization community is beginning to recognize the implications: the VISxGenAI workshop at IEEE VIS~\cite{visxgenai2025} explicitly addresses the question of designing visual tools for and with AI agents.

\subsection{Structured and Accessible Visualization}
A parallel literature explores structured representations that improve machine accessibility to visual data.
DePlot~\cite{liu2023deplot} and MatCha~\cite{liu2023matcha} convert charts to data tables; UniChart~\cite{masry2023unichart} pretrains on chart-table pairs; declarative grammars like Vega-Lite~\cite{satyanarayan2017vegalite} provide machine-parseable specifications.
Bursztyn et~al.~\cite{bursztyn2024representing} find that text-based chart specifications strongly outperform vision-based GPT-4 on explanation generation---evidence that representation format matters substantially for machine performance.
This line of work treats the problem as one of \emph{format translation}: converting visual information into text or structured data that machines can process more easily.

Accessibility research provides a closely related perspective.
Lundgard and Satyanarayan~\cite{lundgard2022accessible} develop a semantic model for natural language descriptions of visualizations; VL2NL~\cite{ko2024vl2nl} translates visualization specifications to natural language.
Zong et~al.~\cite{zong2022rich} design rich screen reader experiences that create representations native to the target modality rather than simply describing existing charts.
Kim et~al.~\cite{kim2021accessible} survey the accessible visualization design space, and Sharif et~al.~\cite{sharif2022voxlens} develop interactive tools for non-visual chart access.
This community has long grappled with how to convey visual information to audiences who process it differently---precisely the challenge machine-oriented design faces.
A recurring finding is that literal translation often fails: Lundgard and Satyanarayan's semantic model shows that effective descriptions require content beyond visual feature enumeration, and Zong et~al.'s work demonstrates the value of representations native to the target modality.
Similarly, machine-oriented visualization may require not pixel-perfect chart optimization but understanding what semantic content machines need and what representational formats they process most effectively.

Our argument departs from these literatures by asking whether the problem admits a complementary response: investigating what visual representations would natively serve machine processing mechanisms.

\subsection{AI-Visualization Pipelines}
Recent work increasingly integrates visualization into AI-mediated analytical workflows.
LIDA~\cite{dibia2023lida} uses LLMs to generate visualizations automatically; DataNarrative~\cite{islam2024datanarrative} combines visualizations with text in automated data storytelling; Data Formulator~\cite{wang2025dataformulator2} enables iterative human-AI visualization creation.
These systems treat visualization primarily as an output artifact for human consumption within an AI-assisted workflow.

An important observation is that these systems already have AI agents \emph{generating} visualizations.
If machine-oriented design principles were to be developed, they could inform not only how machines consume charts but also how AI-generated visualizations are designed when the intended consumer is another machine agent in the pipeline---a scenario that current systems do not explicitly address.
Li et~al.~\cite{li2025visualization} provide initial evidence that visualization can serve as an input that helps AI systems reason about data, though the question of which visual designs are most beneficial remains open.

\section{Visualization's Knowledge Base Is a Science of Human Perception}
\label{sec:knowledge_base}
\subsection{The Historical Contingency of a Human Audience}
The canonical definitions of visualization all center on a human perceiver.
Card, Mackinlay, and Shneiderman~\cite{card1999readings} define visualization as ``the use of computer-supported, interactive, visual representations of data to amplify cognition.''
Munzner~\cite{munzner2014visualization} frames the field as concerned with ``augment[ing] human capabilities rather than replace[ing] people with computational decision-making methods.''
Ware~\cite{ware2012information} grounds the entire enterprise in perception, arguing that effective visualization depends on encoding information in ways aligned with human visual processing capabilities.

These definitions are not arbitrary---but they are \emph{historically contingent}.
They reflect who could use visualization at the time they were written.
Bertin's \emph{Semiology of Graphics} appeared in 1967; Cleveland and McGill ran their psychophysical experiments in 1984; Card et~al.\ synthesized the field in 1999.
Modern vision-language models capable of processing chart images did not exist until the 2020s.
The human-centeredness of visualization is a product of its era, not a logical necessity.

One could, in principle, expand the \emph{definition} of visualization to include machine audiences---just as ``computing'' expanded from human calculators to electronic machines.
But expanding the definition does not expand the \emph{knowledge base}.
Cleveland and McGill's rankings are still about human vision.
Bertin's visual variables still characterize information conveyance to human perceivers.
And therein lies the gap.

\subsection{Every Guideline Is a Claim About Human Perception}
The human-specificity of visualization's knowledge base is not incidental---it is structural.
The field's core contributions are, without exception, claims about how the human visual system processes information:

\begin{itemize}
  \item \textbf{Bertin's visual variables}~\cite{bertin1983semiology}: position, size, shape, value, color, orientation, and texture, characterized by their perceptual properties---selective, associative, quantitative, ordered---based on cartographic practice and theoretical analysis of how graphical marks convey information.
  \item \textbf{Cleveland and McGill's task-based rankings}~\cite{cleveland1984graphical}: position along a common scale is most accurately judged by humans, followed by position on non-aligned scales. Length, direction, and angle share the next tier of accuracy. Area, volume, and curvature follow, with color saturation and density least accurately judged. These rankings derive from perceptual experiments measuring human judgment accuracy, and have been replicated and extended through crowdsourced experiments~\cite{heer2010crowdsourcing,kim2018assessing}, always measuring \emph{human} performance.
  \item \textbf{Preattentive processing}~\cite{healey2012attention}: certain visual features---color, orientation, size, motion---are detected automatically by the human visual cortex in under 200ms, enabling rapid identification of outliers and patterns.
  \item \textbf{Gestalt principles}: proximity, similarity, closure, continuity---these describe how \emph{human brains} organize visual information into coherent groups and patterns.
  \item \textbf{Color science}: perceptually uniform color spaces (CIE~Lab), colorblind-safe palettes, and luminance contrast guidelines all derive from the physiology of human cone cells and the neuroscience of human color perception.
\end{itemize}

Tufte's data-ink ratio~\cite{tufte2001visual}, Wilkinson's grammar of graphics~\cite{wilkinson2005grammar}, and Munzner's nested model~\cite{munzner2014visualization} all presuppose a human viewer.
This is not a limitation of these contributions---they are rigorous, empirically validated, and enormously valuable.
But their scope is specific: they describe design principles for a particular perceptual system.

\subsection{The Scope Problem}
The implication is that expanding visualization's audience to include machines creates a gap in the knowledge base.
The field possesses extensive, well-validated design principles for one audience (humans) and has not yet investigated what design principles might serve the other (machines).

Distributed cognition theory~\cite{hutchins1995cognition} clarifies why this gap is fundamental rather than incidental.
Kirsh~\cite{kirsh1995intelligent} demonstrates that effective external representations reduce memory load and computational steps by exploiting \emph{specific} cognitive constraints---limited working memory, sequential attention, and perceptual grouping mechanisms.
Zhang and Norman~\cite{zhang1994representations} formalize how the distribution of information across internal and external representations determines problem-solving efficiency, with optimal external representations depending critically on the architecture of the internal system.
This framework suggests that human visualization design is not merely \emph{optimized} for human perception---it is \emph{constituted} by assumptions about human cognitive architecture.
Gestalt grouping assumes a visual system that segments by proximity and similarity; preattentive feature recommendations assume parallel feature maps with capacity limits; color encoding guidelines assume trichromatic vision with specific opponent channels.
Machines have different constraints: large but finite context windows rather than limited working memory, learned rather than evolved feature channels, and attention mechanisms architecturally different from human sequential fixation.
The representations that scaffold human cognition may provide no scaffolding---or different scaffolding---for artificial systems.

An important clarification is necessary here.
VLM image processing is not ``perception'' in the biological sense---it is learned statistical decoding of pixel patterns, shaped by training data, architecture choices, and optimization objectives.
Unlike human vision, which is a stable evolved system with broadly consistent properties across individuals, machine visual processing is engineered, rapidly evolving, and potentially unstable across architectures.
But this distinction actually \emph{strengthens} the case for investigating new design foundations.
Precisely because machine visual processing is fundamentally different in mechanism from human perception---not evolved but designed, not stable but shifting---there is no reason to expect that design principles derived from human psychophysics would transfer.
One could attempt to engineer VLMs to work optimally with human-designed charts, but this would amount to constraining machine cognition to human design conventions rather than exploring what visual representations might best serve these different information-processing systems.

As AI agents are increasingly embedded in analytical workflows---reading dashboards, interpreting automated reports, processing visual data summaries within multi-agent pipelines~\cite{islam2024datanarrative,dibia2023lida,wang2025dataformulator2}---the absence of investigation into machine-oriented design means these agents receive visual representations designed entirely for a different consumer.
The field has not yet systematically asked which encodings machines extract most reliably, which layouts facilitate machine comprehension, or which design patterns help or hinder machine processing.

Lundgard and Satyanarayan's four-level semantic model~\cite{lundgard2022accessible} illuminates this point.
Their framework distinguishes perceiver-independent content (Levels~1--2: chart construction properties, statistical features) from perceiver-dependent content (Levels~3--4: perceptual patterns, domain-specific insights).
What a perceiver extracts from a visualization depends on the perceiver.
Humans excel at extracting Level~3--4 content---recognizing trends, spotting anomalies, perceiving clusters---through Gestalt grouping and preattentive processing.
Machines, as we discuss in the next section, appear to rely more heavily on Level~1--2 extraction and struggle precisely where human perception is strongest.

This asymmetry points to a methodological need.
The field's primary method for developing design knowledge---controlled experiments with human subjects~\cite{cleveland1984graphical,heer2010crowdsourcing,kim2018assessing}---is inherently calibrated to human responses.
Poonam et~al.~\cite{poonam2025evaluating} have begun adapting this methodology, replicating Cleveland and McGill's experiments with Vision Transformers.
The broader research program would extend this approach: systematically testing which visual designs enable better machine performance, using the benchmark infrastructure that already exists but reframing it from model evaluation to design evaluation.
Benchmarks currently test models; they could also test designs.

\section{Machines See Charts Differently}
\label{sec:machines_see_differently}
\subsection{The Gap Is Qualitative, Not Just Quantitative}
A large and rapidly growing benchmark literature documents VLM performance on chart understanding tasks.
The headline numbers are substantial: CharXiv~\cite{wang2024charxiv} reports a 33-point gap between human performance (80.5\%) and GPT-4o (47.1\%) on realistic chart comprehension; ChartMuseum~\cite{tang2025chartmuseum} finds 93\% human accuracy versus 63\% for the best model; ChartInsights~\cite{wu2024chartinsights} reports that 19 multimodal large language models average only 39.8\% on low-level chart tasks.
Bendeck and Stasko~\cite{bendeck2025empirical} find GPT-4 scoring at the 16th percentile on the VLAT visualization literacy assessment.

But the more revealing finding is not the size of the gap---it is its \emph{character}.
Bendeck and Stasko~\cite{bendeck2025empirical} observe that GPT-4 demonstrates understanding of trends and design best practices but struggles with simple value retrieval---the \emph{opposite} pattern from humans, who are strong at value retrieval but sometimes weaker on trend synthesis.
Verma et~al.~\cite{verma2025chart6} evaluate eight VLMs across six visualization literacy assessments and find that all models produce error patterns reliably distinct from those of human participants---not just lower accuracy, but different patterns of success and failure.
Machine performance is also \emph{brittle}: CharXiv documents that open-source models exhibit up to 34.5\% performance drops on stress tests with slight chart modifications that do not affect human accuracy, with proprietary models showing smaller but still significant degradation~\cite{wang2024charxiv}; ChartMuseum finds that human performance remains stable as visual reasoning demands increase while model performance degrades significantly~\cite{tang2025chartmuseum}.
This pattern of divergence---not uniform weakness but qualitatively different processing---is what motivates the question of whether different design foundations are needed.

\subsection{The Machine Encoding Hierarchy}
\label{sec:machine_hierarchy}
Diagnostic evidence comes from EncQA~\cite{mukherjee2025encqa}, which systematically tests VLM performance across six visual encoding channels and eight chart understanding tasks.
The results reveal encoding-dependent performance patterns, though the authors caution that ``performance varies significantly across encodings within the same task, as well as across tasks'' and that ``the same ranking of encodings might not apply for all tasks.''
With these caveats, some general patterns emerge:

\begin{itemize}
  \item \textbf{Higher accuracy}: Position aligned to axes---where values can be read directly from axis gridlines.
  \item \textbf{Moderate accuracy}: Length, nominal color, shape---encodings requiring legend interpretation or spatial comparison.
  \item \textbf{Lower accuracy}: Area, color quantitative (lightness)---encodings requiring legend-based estimation rather than direct axis reading.
\end{itemize}

These patterns may reflect the beginnings of a machine encoding hierarchy distinct from Cleveland and McGill's human perceptual rankings, but should be understood as task-dependent tendencies in current VLMs rather than a stable ordering established with comparable rigor.

Poonam et~al.~\cite{poonam2025evaluating} provide independent corroboration by replicating Cleveland and McGill's foundational experiments with Vision Transformers, finding that the resulting effectiveness rankings differ from human rankings.
Pandey and Ottley~\cite{pandey2025benchmarking} report consistent patterns: on bubble charts, which rely on area encoding, accuracy ranged from 18.6\% to 61.4\% across models.

We should be candid about the limitations of this evidence.
These patterns reflect current architectures and may shift as VLMs evolve---higher-resolution vision encoders, different patch sizes, or fundamentally different processing mechanisms could alter the ordering.
Model size does not improve performance for many task-encoding pairs in the current EncQA data~\cite{mukherjee2025encqa}, but this finding comes from one generation of models and should not be overinterpreted as permanent.
What the evidence does establish is that the current encoding performance patterns are \emph{different from} the human ones---not simply a degraded version of them---and that this difference warrants investigation.

\subsection{How VLMs Actually Process Charts}
The architectural basis for these differences is at least partially understood.
Vision-language models process chart images through mechanisms fundamentally unlike human vision:

\textbf{Patch-based tokenization.}
Following the Vision Transformer (ViT) architecture~\cite{dosovitskiy2021image}, VLMs divide input images into fixed-size patches---typically $16 \times 16$ pixels---each treated as a token and processed through transformer layers.
Visual elements are decomposed into a grid, not perceived holistically as in human vision.
A bar in a bar chart is not a unified perceptual object but a collection of patches that must be integrated through attention.
This decomposition has direct consequences for chart understanding: a legend in the top-right corner and the data marks it explains in the center are separated by many patches, requiring long-range attention to connect them.
Humans perform this integration effortlessly through Gestalt principles of similarity; machines must learn it from data.

\textbf{Texture bias.}
Geirhos et~al.~\cite{geirhos2019imagenet} demonstrated that convolutional neural networks exhibit a \emph{texture bias} where humans exhibit a \emph{shape bias}.
Models attend to surface-level patterns---textures, local statistics---rather than the structural forms that human perception privileges.
In chart understanding, this manifests in multiple ways.
Machines may attend to the texture of a filled bar rather than its height; they may be influenced by background patterns or grid densities that humans filter automatically; and they may confuse charts that look superficially similar (e.g., similar color palettes) but represent different data.
The adversarial vulnerability of neural networks to imperceptible perturbations~\cite{szegedy2014intriguing} is a related manifestation: small changes to pixel values that are invisible to humans can dramatically alter model outputs.

\textbf{OCR-reliance.}
VLMs depend heavily on text extraction from chart images.
Much of what VLMs ``see'' in charts is text they read, not visual patterns they perceive---the inverse of human chart reading, where visual encoding carries the primary information and text labels serve as confirmation.
This reliance on OCR has a further implication: the legibility and placement of text within a chart image is a first-order design concern for machine audiences, not a secondary aesthetic consideration.
Text that is too small for reliable OCR extraction, or that overlaps with data marks, directly impairs machine comprehension in ways that it does not for humans (who can usually resolve such ambiguities through context).

\textbf{Sequential versus parallel processing.}
Human vision operates in parallel across the visual field: preattentive features (color, orientation, size) are detected simultaneously across the entire scene~\cite{healey2012attention}, enabling the rapid ``pop-out'' effect that makes outliers and patterns immediately salient.
Transformer attention, by contrast, is fundamentally sequential: the model attends to different regions in sequence (even if parallelized computationally), and the order and scope of attention are learned rather than hardwired.
This means that visual designs relying on the human ``pop-out'' effect---a single red point among blue points, a single tall bar among short bars---may not be equivalently effective for machines, which must learn to attend to the salient region rather than detecting it automatically.

\subsection{Machine-Specific Failure Modes}
The benchmark literature reveals failure modes qualitatively different from human errors:

\textbf{Deception vulnerability.}
Mahbub et~al.~\cite{mahbub2025perils} tested 10 VLMs on charts with misleading design elements.
All 10 models were vulnerable: inverted axes affected most or all models tested, though Gemini-2.5-Pro showed some ability to detect the manipulation; truncated axes affected 7 of 10, and aspect ratio distortion misled 9 of 10.
Pandey and Ottley~\cite{pandey2025benchmarking} confirmed this at scale: all VLMs tested scored at or below 30\% on detecting misleading visualization elements.
This finding is diagnostically interesting---it reveals something about how machine vision processes structural chart properties---though its design implications are uncertain, since what constitutes ``deception'' for a machine may differ from what misleads a human.

\textbf{Overconfident hallucination.}
CHART~NOISe~\cite{shin2025chartnoise} documents that VLMs under visual degradation exhibit value fabrication, trend misinterpretation, and entity confusion---while maintaining high confidence in their incorrect answers.
Unlike human readers, who typically express uncertainty when visual information is ambiguous, models produce precise but fabricated numerical values.

\textbf{Complexity degradation.}
ChartX~\cite{xia2024chartx} identifies the chart types most difficult for AI---rose charts, 3D bar charts, bubble charts, multi-axis charts, radar charts, and area charts---all requiring spatial reasoning beyond simple position reading.
The pattern is consistent: as visual complexity increases, machine performance degrades more sharply than human performance.

\textbf{An honest assessment.}
The current benchmark evidence documents more machine failures than successes on chart understanding---but this observation requires careful interpretation.
The benchmarks themselves are designed around tasks that human perception handles well: value retrieval, legend lookup, spatial comparison.
Measuring machines against human-oriented tasks on human-designed charts, and concluding they are ``weaker,'' is circular if the argument is that machines need different designs.

Three findings illustrate why the framing should be \emph{different} rather than \emph{weaker}.
First, Bendeck and Stasko~\cite{bendeck2025empirical} find that GPT-4 shows strength on trend understanding while struggling with simple value retrieval---the opposite of human patterns, not a uniform degradation.
Second, Neo et~al.~\cite{neo2025interpreting} show that VLMs exhibit genuine spatial processing: object information is localized to visual tokens corresponding to spatial positions, and ablating object-specific tokens causes over 70\% accuracy drops.
Third, Schulze Buschoff et~al.~\cite{schulzebuschoff2025visual} find that GPT-4 and Claude-3 perform above chance on visual cognition tasks requiring genuine visual understanding, with more capable models showing more robust performance.

The honest summary is that machines process charts \emph{differently}---through different mechanisms, with different strengths and different failure modes.
How they would perform on visual representations designed for their processing characteristics is unknown, because such representations do not yet exist.
That unknown is precisely the research gap this paper identifies.

\subsection{Summary: Two Perceptual Systems Compared}
Table \ref{tab:perceptual_comparison} summarizes the key differences between human and machine visual processing as they relate to chart understanding.
The pattern of differences---different mechanisms, different strengths, different failure modes---suggests these are not simply differences of degree.
Whether these differences are stable enough to ground design principles, or whether they will shift substantially as architectures evolve, is an open question---but the divergence from human processing is clear enough to motivate investigation.

\begin{table*}[!t]
  \caption{%
    Comparison of human and machine visual processing for chart understanding, synthesized from benchmark evidence.
    The pattern suggests qualitatively different processing mechanisms, not merely lower machine accuracy.%
  }
  \label{tab:perceptual_comparison}
  \centering
  \scriptsize
  \begin{tabular}{p{2.8cm}p{5.8cm}p{5.8cm}}
    \toprule
    \textbf{Dimension} & \textbf{Human Perception} & \textbf{Machine Processing (Current VLMs)} \\
    \midrule
    Processing mechanism & Holistic scene perception via retinal $\rightarrow$ cortical pathways; parallel feature extraction & Patch-based tokenization ($16 \times 16$ px); sequential attention integration~\cite{dosovitskiy2021image} \\
    \addlinespace
    Encoding hierarchy & Position (aligned) $>$ position (non-aligned) $>$ length, direction, angle $>$ area, volume, curvature $>$ color saturation, density~\cite{cleveland1984graphical} & Task-dependent; axis-aligned encodings generally outperform legend-based encodings, but rankings vary by task~\cite{mukherjee2025encqa} \\
    \addlinespace
    Primary information channel & Visual encoding (shape, position, color); text confirms & Text (OCR extraction); visual encoding supplements~\cite{wu2024chartinsights} \\
    \addlinespace
    Shape vs.\ texture & Shape bias---structural form drives recognition & Texture bias---surface statistics drive recognition~\cite{geirhos2019imagenet} \\
    \addlinespace
    Robustness & Robust to minor perturbations, color changes, noise & Brittle---up to 34.5\% drop from minor modifications (open-source models); proprietary models also affected~\cite{wang2024charxiv} \\
    \addlinespace
    Complexity handling & Gestalt grouping and preattentive processing maintain performance under high complexity & Performance degrades as complexity increases~\cite{tang2025chartmuseum,xia2024chartx} \\
    \addlinespace
    Response to misleading designs & Can often identify and compensate for truncated axes, inverted scales, etc. & Vulnerable to human-defined misleading patterns ($\leq$30\% detection); whether machines have distinct failure modes is an open question~\cite{mahbub2025perils,pandey2025benchmarking} \\
    \addlinespace
    Legend interpretation & Fluent---color/shape legends parsed automatically & Difficult---requires cross-image spatial reasoning~\cite{mukherjee2025encqa} \\
    \addlinespace
    Confidence calibration & Expresses uncertainty when information is ambiguous & Overconfident hallucination---fabricates precise values~\cite{shin2025chartnoise} \\
    \addlinespace
    Spatial integration & Effortless across entire visual field & Limited by patch boundaries and attention span~\cite{dosovitskiy2021image} \\
    \bottomrule
  \end{tabular}
\end{table*}

\section{The Limits of Bypassing Vision}
\label{sec:against_bypassing}
Given the evidence that VLMs struggle with human-designed charts, the prevailing research response has been to route around machine vision entirely.
MatCha~\cite{liu2023matcha} introduces a combined pretraining approach including ``chart derendering''---converting chart images back to data tables---alongside math reasoning pretraining, achieving up to 20\% improvement in question-answering accuracy.
DePlot~\cite{liu2023deplot} translates plots to tables and achieves 29.4\% improvement over fine-tuned state-of-the-art models.
Structured formats---Vega-Lite JSON~\cite{satyanarayan2017vegalite}, semantic SVG, data tables---are consistently shown to improve AI parsing compared to raster chart images~\cite{masry2023unichart}.
Bursztyn et~al.~\cite{bursztyn2024representing} find that text-based chart specifications strongly outperform vision-based GPT-4 on explanation generation.

These results are real and pragmatically important.
We do not argue that bypassing vision is wrong---in many current pipelines it is the most effective approach.
We argue that it forecloses a question worth investigating: whether visual representations designed for machine cognition could be valuable.

\subsection{The Distinction Worth Preserving}
The derendering evidence demonstrates that charts designed according to \emph{human} perceptual principles are obstacles for machines.
It does not demonstrate that \emph{all possible visual representations} are obstacles for machines.
These are different claims.
The first motivates investigating alternative visual representations for machine audiences; the second motivates abandoning visual representation altogether.
The benchmark literature supports the first claim.
Li et~al.'s~\cite{li2025visualization} finding that scatterplots help AI data analysis, while a single study that should not be overinterpreted, provides initial evidence against the second.

\subsection{The Open Question of Spatial Representation}
For human viewers, visualization's core value proposition is \emph{spatial data summarization}: compressing data points into spatial patterns---clusters, trends, outliers, distributions.
Whether an analogous value proposition exists for machines is an open question that the field has not yet addressed.

We should be honest about what we do and do not know here.
We do not know whether ``spatial summarization'' is the right concept for machine visual processing.
Machines do not perceive spatial patterns the way humans do---they process patches through attention, not scenes through Gestalt grouping.
The mechanism by which visual representation could benefit machines may be something different entirely: redundant encoding, a different compression of information, or a form of scaffolding we have not yet characterized.

What we do know is suggestive.
Wu et~al.~\cite{wu2024vot} find that prompting LLMs to generate visual reasoning traces improves spatial reasoning by up to 23.5 percentage points over baselines without visualization---even when the model can reason textually.
Hu et~al.~\cite{hu2024sketchpad} show that giving multimodal models a ``sketchpad'' to draw spatial artifacts during reasoning yields 12.7\% gains on math tasks.
Li et~al.~\cite{li2025visualization} find that adding scatterplots to raw data improves analysis even when all data is provided.
These findings do not prove that machine-oriented visualization would be valuable---but they suggest that visual representation provides \emph{something} to machine cognition beyond what text and data alone provide, and that this something is worth understanding.

\subsection{Pragmatic Objections}
Three pragmatic objections deserve honest engagement.

\textbf{Code execution.}
Modern AI agents can execute code: an agent receiving a table can compute correlations, run clustering, or fit regressions with perfect numerical precision.
Why would spatial representation benefit a system that can explicitly compute patterns?
This is a strong objection.
However, humans \emph{can} also compute correlations from tables, yet visualization still provides value as cognitive scaffolding.
The Wu et~al.\ and Hu et~al.\ findings above suggest---though do not prove---that an analogous scaffolding effect may exist for machines.

\textbf{Source data availability.}
In most pipelines, charts are generated \emph{from} structured data; the source table already exists.
Why render pixels to feed a vision encoder when the data is available directly?
This is perhaps the strongest pragmatic objection.
We acknowledge it fully: in pipelines where source data is available, bypassing vision may often be the most efficient choice.
Our argument is not that machines should always receive visual representations instead of data.
It is that the field has not yet investigated whether visual representations designed for machine cognition could add value---even in the presence of source data, as Li et~al.'s findings tentatively suggest---and that this investigation is scientifically worthwhile.

\textbf{Self-directed visualization.}
AI agents can generate their own visualizations: systems like LIDA~\cite{dibia2023lida} already have agents creating charts as part of analytical workflows.
An agent could design a visualization tailored to its specific query rather than receiving a pre-rendered chart.
This objection is well-taken---and we view it as \emph{supporting} rather than undermining the research direction.
If agents will generate visualizations for their own consumption, then understanding what visual designs serve machine cognition becomes even more important.
Currently, when an AI agent generates a chart, it uses human-oriented design defaults (color legends, Gestalt-dependent layouts, aesthetic conventions).
If machine-oriented design foundations existed, self-generated visualizations could apply them---a scenario where the design knowledge this paper calls for would be directly actionable.

\subsection{What Might Be Lost by Not Investigating}
We are not arguing that bypassing vision is wrong.
We are arguing that exclusively bypassing vision---treating it as the settled answer---means the field never investigates an alternative that could be valuable.
If it turns out that visual representations designed for machine cognition provide no benefit beyond structured data, that is a useful empirical finding.
If it turns out they do provide benefit---through spatial scaffolding, redundant encoding, or mechanisms we have not yet identified---then the field will have missed an important research direction by assuming the answer in advance.

The cost of investigating is modest: the experimental infrastructure exists in the VLM benchmark literature.
The cost of not investigating is that we foreclose the possibility that visualization's value proposition extends, in some form, beyond its original human audience.

\section{Machine Visual Processing Is Partially Understood}
\label{sec:machine_processing}
The preceding sections documented \emph{that} machines process charts differently from humans.
This section asks what follows for design---and argues that, despite the temptation, prescribing specific design principles would be premature.

\subsection{The Design-Relevance of Current Observations}
Section \ref{sec:machines_see_differently} established several properties of VLM visual processing: patch-based decomposition, texture bias, OCR-reliance, sequential attention, and different encoding performance patterns.
Each is potentially design-relevant, but the relationship between architectural property and design implication is less direct than it may appear.

Consider the most robust current finding: text reliance.
It is tempting to derive a design principle from it (``add more text labels'').
But if machine-oriented design were simply ``add text to everything,'' it would reduce to formatted text tables with spatial positioning---and the question of whether that constitutes meaningful ``visualization'' would be fair.
The more interesting question is whether future machine cognition will develop stronger genuinely visual processing, and what design foundations would serve that development.

Similarly, EncQA~\cite{mukherjee2025encqa} and Poonam et~al.~\cite{poonam2025evaluating} find that position aligned to axes is currently the most reliably processed encoding for machines---but whether this reflects a deep architectural property or a training artifact is unclear.
Machine brittleness under perturbations~\cite{wang2024charxiv} and complexity~\cite{tang2025chartmuseum} may likewise be a property of current training rather than an inherent limitation.

At the same time, the evidence is not uniformly negative about machine visual capabilities.
Neo et~al.~\cite{neo2025interpreting} show genuine spatial processing in VLMs: object information is localized to visual tokens at corresponding spatial positions.
Shtedritski et~al.~\cite{shtedritski2023redcircle} find that CLIP responds to geometric visual cues.
Yang et~al.~\cite{yang2023setofmark} show that spatial markers dramatically improve VLM visual grounding.
Characterizing current VLMs as ``OCR plus weak vision'' understates capabilities that exist but are not well understood.

The honest assessment is that the current evidence is sufficient to establish that machine visual processing \emph{differs} from human visual processing in design-relevant ways---but insufficient to determine what the optimal design response would be.

\subsection{Why We Do Not Prescribe Design Principles}
It would be straightforward to derive a set of design principles from the observations above: make marks patch-coherent, anchor data to axes, integrate text labels, reduce clutter, add redundant encoding.
We believe this would be a mistake, for two reasons.

First, machine cognition is evolving rapidly.
Design principles derived from current VLM characteristics risk becoming obsolete with the next generation of vision encoders.
Higher-resolution inputs, different patch sizes, native multi-image support, and fundamentally different architectures could shift encoding performance patterns substantially.
The field of human-oriented visualization built its design knowledge on a stable biological substrate---human vision changes on evolutionary timescales.
Machine vision changes on engineering timescales.
Prescribing specific design principles today would be premature.

Second, and more fundamentally, we do not know what machine-oriented visualization \emph{should} look like.
The optimal form might bear no resemblance to conventional charts---it could exploit machine processing mechanisms in ways we cannot currently anticipate.
Constraining the design space to variations on human chart types (bar charts with more labels, scatterplots with less clutter) may miss the most valuable possibilities.
The research program should explore the full design space, not anchor prematurely to familiar forms.

What we \emph{can} prescribe is the \emph{methodology}: the field should investigate machine-oriented visual design through systematic empirical experiments, the way Cleveland and McGill investigated human-oriented design.
The contribution is identifying the need for this investigation, not filling it.

\subsection{The Need for a Machine Bertin}
\label{sec:machine_bertin}
Bertin's \emph{Semiology of Graphics}~\cite{bertin1983semiology} established visual variables and characterized their perceptual properties based on cartographic practice and theoretical analysis of human visual processing.
Cleveland and McGill~\cite{cleveland1984graphical} validated and extended this through controlled perceptual experiments.
Heer and Bostock~\cite{heer2010crowdsourcing} replicated these rankings through crowdsourced studies.

The field needs to begin a comparable investigation for machine visual processing---a ``machine Bertin.''
EncQA~\cite{mukherjee2025encqa} and Poonam et~al.~\cite{poonam2025evaluating} have taken the first steps, establishing initial encoding performance patterns for VLMs.
But these are fragments, not a theory.
Extending this to more encoding channels, data types, model architectures, analytical tasks, and---crucially---to visual representations that go beyond conventional chart types is the central empirical challenge.

The experimental infrastructure exists: the VLM benchmark literature already runs the kind of controlled evaluations needed.
What is missing is the \emph{design-oriented framing} that asks not ``how well does this model perform on this chart?'' but ``which visual designs enable better machine performance, and why?''
The shift from model evaluation to design evaluation is the methodological contribution this paper calls for.

\section{Human-Oriented and Machine-Oriented Visualization}
\label{sec:two_layer}
\subsection{The Core Argument}
The evidence reviewed in the preceding sections suggests a conceptual distinction worth making explicit (Figure \ref{fig:teaser}):

\begin{enumerate}
  \item \textbf{Human-oriented visualization:} Visualizations designed for human perception following established design principles---Cleveland and McGill, Bertin, preattentive processing, Gestalt grouping, perceptually uniform color.
  This is the knowledge base the field already possesses.
  \item \textbf{Machine-oriented visualization:} Visual representations designed for machine cognition, following design principles that do not yet exist but that the field could develop through systematic empirical investigation.
\end{enumerate}

From an information-theoretic perspective, visualization is a communication channel optimized for human perceptual bandwidth.
Humans have specific channel characteristics: limited conscious throughput, parallel preattentive processing of specific features, and sequential attention for conjunction search.
Visualization design exploits these characteristics---using preattentive features for efficient parallel encoding, limiting simultaneous encodings to avoid conjunction search, and leveraging pattern recognition for compression.
Machines have radically different channel characteristics: high bandwidth for structured data, no preattentive/attentive distinction, and different bottlenecks (context windows, attention heads).
An encoding optimized for one receiver may be suboptimal for another, even when both can eventually decode the message.

This is a \emph{conceptual} distinction, not an engineering architecture.
We are not proposing that every pipeline must maintain two rendering paths, or that specific systems should be redesigned.
We are arguing that these are different design problems requiring different knowledge bases---and that the second knowledge base does not yet exist.

Human-oriented and machine-oriented visualizations need not coexist in the same system.
A dashboard designed for human analysts is one design problem.
A visual representation generated by an AI agent for its own reasoning---or consumed by another agent in a pipeline---is a different design problem.
The point is that the design foundations for the second problem have not been investigated, and the evidence suggests they should be.

\subsection{Why This Is Not Format Translation}
The distinction we draw is between designing for a different consumer and \emph{translating} an existing design into a different format.
Converting a human-designed chart to a data table (DePlot~\cite{liu2023deplot}), to structured text (UniChart~\cite{masry2023unichart}), or to a specification language (Vega-Lite~\cite{satyanarayan2017vegalite}) are all forms of format translation: taking a representation designed for human perception and converting it to something machines process more easily.

The alternative we are proposing the field investigate is \emph{native design}: creating visual representations whose design foundations are derived from how machines process visual information.
The distinction matters because format translation is constrained by the original human-oriented design---it can only convert what was already there.
Native design could explore representations that no human designer would create, because they are not optimized for human perception.

We should be clear about how uncertain this territory is.
We do not know whether native machine-oriented visual design would outperform format translation, structured data, or code execution for any given task.
We do not know what such designs would look like---they might be completely alien relative to current chart types.
Whether this direction proves fruitful is an empirical question that requires the investigation we call for in Section \ref{sec:agenda}.

\subsection{The Architecture Dependence Problem}
A natural concern is that design principles for machines would be architecture-specific: what works for GPT-4o might not work for Gemini or for future architectures.
This is a legitimate concern, and we take it seriously.

We are currently in early stages of VLM adoption, and cross-model variation in chart processing is real~\cite{mukherjee2025encqa,verma2025chart6}.
However, technology adoption tends toward convergence as platforms standardize---as happened with web browsers, mobile operating systems, and display technologies.
Whether VLM visual processing will similarly converge is an open empirical question.
Chen and Bonner~\cite{chen2024universal} find that diverse neural network architectures converge on a shared set of universal visual dimensions, and Kazemian et~al.~\cite{kazemian2025cortex} show that certain visual processing properties are intrinsic to architectural constraints rather than learned.
These findings are suggestive but come from general visual processing, not chart understanding specifically.
Whether chart-relevant processing properties converge across architectures is a question the research program would need to answer.

Even if architecture-dependent variation turns out to be significant, identifying that human design principles do not transfer to machines is valuable.
The field would know that machine-oriented design is needed, even if the specific principles are model-dependent---just as responsive web design adapts to different screen sizes while still being informed by a shared understanding of layout principles.

\subsection{Relationship to Existing Technologies}
Declarative grammars like Vega-Lite~\cite{satyanarayan2017vegalite} and structured visual formats (SVG, semantic markup) occupy an important middle ground.
They preserve spatial relationships in machine-parseable form and may prove to be a practical delivery format for machine-oriented representations.
We view these technologies as complementary: they could serve as authoring or delivery mechanisms for whatever machine-oriented designs the research program develops.

The key claim is not about delivery format---pixels, SVG, or something else---but about \emph{design foundations}.
Machine audiences may need representations designed for their processing characteristics, regardless of the format in which those representations are delivered.

\section{Research Agenda}
\label{sec:agenda}
If the field accepts that machine-oriented visual design is worth investigating, what are the key research questions?
We outline five directions, ordered from foundational to applied.

\textbf{1. Empirical foundations: the machine Cleveland and McGill.}
The field needs systematic experiments on VLMs analogous to Cleveland and McGill's ranking studies for humans.
Which visual encodings do machines extract most accurately, across what data types and tasks?
How do layout, aspect ratio, label placement, and color choices affect machine comprehension?
EncQA~\cite{mukherjee2025encqa} and Poonam et~al.~\cite{poonam2025evaluating} provide models for this methodology; extending it comprehensively is a high-priority research need.
Critically, these experiments should not be limited to existing chart types---the design space should include unconventional visual representations that may serve machine processing in ways no human-oriented design does.
Concretely, three methodological paradigms are needed: (a)~\emph{encoding effectiveness rankings}---replicating Cleveland and McGill's paradigm with VLMs, systematically varying visual encoding (position, length, angle, area, color, shape) while holding data constant to determine whether machines show the same or different hierarchies; (b)~\emph{efficiency metrics beyond accuracy}---machine-oriented evaluation should include token usage, inference cost, and robustness to styling variations, since a visualization requiring elaborate chain-of-thought prompting may be less ``effective'' than one parsed immediately, even at equal final accuracy; and (c)~\emph{format comparison studies}---presenting identical data as chart images, markdown tables, JSON, and natural language descriptions, measuring comprehension across task types to reveal when visual encoding helps versus hurts machine understanding.

\textbf{2. Benchmark reorientation: from model evaluation to design evaluation.}
Existing benchmarks~\cite{masry2022chartqa,wang2024charxiv,tang2025chartmuseum} evaluate model capabilities on fixed charts.
The field needs benchmarks that evaluate \emph{design choices}---asking not ``how well does this model read this chart?'' but ``does this model extract values more accurately from design A or design B of the same data?''
This reframing---from testing models to testing designs---is a modest methodological shift but a significant conceptual one.

\textbf{3. Cross-architecture generalization.}
How much do machine-oriented design findings generalize across architectures?
If design principles derived from GPT-4o's vision encoder do not transfer to Gemini or future architectures, the research program risks reducing to model-specific optimization.
Mapping the degree of cross-architecture convergence for chart-specific processing is essential for determining whether generalizable design principles are possible.
The convergence evidence from general visual processing~\cite{chen2024universal,kazemian2025cortex} is suggestive but not conclusive for charts specifically.

\textbf{4. The mechanism question.}
What do machines gain from visual representation, if anything, beyond what structured data provides?
Is it spatial summarization in some form, redundant encoding, a different compression of information, or something else entirely?
Li et~al.~\cite{li2025visualization} show that scatterplots help machine data analysis even when raw data is provided, and Wu et~al.~\cite{wu2024vot} show that visual reasoning traces improve spatial reasoning---but the mechanism is not understood.
Understanding \emph{why} visual representation can benefit machine cognition is essential for designing representations that amplify this benefit.

\textbf{5. Self-directed machine visualization.}
AI agents already generate their own visualizations through systems like LIDA~\cite{dibia2023lida} and Data Formulator~\cite{wang2025dataformulator2}.
When an agent generates a chart for its own consumption, what design should it use?
Currently these systems apply human-oriented defaults.
Investigating whether machine-oriented designs improve agent self-directed reasoning is a natural and practically relevant test of the research direction.

Table \ref{tab:agenda} summarizes these five directions, the core question each addresses, the existing work that provides a starting point, and the primary gap that remains.

\begin{table*}[!t]
  \caption{%
    Summary of the proposed research agenda for machine-oriented visual design.
    Each direction builds on existing work but requires a reorientation from model evaluation to design evaluation.%
  }
  \label{tab:agenda}
  \centering
  \scriptsize
  \begin{tabular}{p{2.6cm}p{3.4cm}p{3.8cm}p{3.8cm}}
    \toprule
    \textbf{Research Direction} & \textbf{Core Question} & \textbf{Existing Starting Points} & \textbf{Primary Gap} \\
    \midrule
    1.\ Empirical foundations & Which visual encodings do machines extract most accurately, and under what conditions? & EncQA~\cite{mukherjee2025encqa}; Poonam et~al.~\cite{poonam2025evaluating} replicate Cleveland \& McGill for VLMs & Systematic coverage of encoding channels, data types, and unconventional representations beyond existing chart forms \\
    \addlinespace
    2.\ Benchmark reorientation & Does design A or design B better serve machine comprehension of the same data? & ChartQA~\cite{masry2022chartqa}; CharXiv~\cite{wang2024charxiv}; ChartMuseum~\cite{tang2025chartmuseum} & Benchmarks evaluate models on fixed charts; none vary design while holding data constant \\
    \addlinespace
    3.\ Cross-architecture generalization & Do machine-oriented design findings transfer across VLM architectures? & Convergence evidence for general visual processing~\cite{chen2024universal,kazemian2025cortex} & No chart-specific cross-architecture studies exist \\
    \addlinespace
    4.\ The mechanism question & Why does visual representation benefit machine cognition beyond structured data? & Li et~al.~\cite{li2025visualization}; Wu et~al.~\cite{wu2024vot}; Hu et~al.~\cite{hu2024sketchpad} & Benefit is documented but mechanism is not understood \\
    \addlinespace
    5.\ Self-directed visualization & What designs should AI agents use when generating charts for their own reasoning? & LIDA~\cite{dibia2023lida}; Data Formulator~\cite{wang2025dataformulator2} & Systems apply human-oriented defaults; no investigation of machine-oriented alternatives \\
    \bottomrule
  \end{tabular}
\end{table*}

\subsection{Open Questions}
Several questions are genuinely open and represent productive research directions.

\textbf{What does machine-oriented visualization look like?}
We deliberately avoid speculating.
Current evidence would suggest text-heavy, simplified charts---but this may reflect the immaturity of current VLMs rather than a fundamental property.
As machine vision capabilities mature, the optimal visual representation for machines may diverge from familiar chart forms in ways that are difficult to predict from current evidence alone.

\textbf{Is ``spatial summarization'' the right concept?}
Visualization's value for humans is often described in terms of spatial data summarization---compressing data into perceivable spatial patterns.
Whether machines benefit from spatial representation in an analogous way, or through a different mechanism entirely, is unknown.
The concept may need to be rethought for machine audiences rather than assumed to transfer.

\textbf{Dual-purpose design.}
Can a single visual design serve both human and machine audiences well, or are the processing differences too large for dual-purpose representations?
Some overlap exists (both benefit from position-based encoding and clear labels), but the tensions may be significant.
Identifying the set of designs that are effective for both audiences---if it exists---is an empirical question.

\textbf{Evaluation metrics.}
How should machine-oriented visualizations be evaluated?
Human-oriented visualization uses task completion time, accuracy, and subjective satisfaction.
Machine-oriented visualization would need analogous metrics: extraction accuracy, downstream reasoning quality, and robustness.
Developing standardized evaluation protocols is essential for the research program to progress.

\textbf{The ``bitter lesson'' objection.}
One might object that investigating machine-oriented design principles contradicts the ``bitter lesson'' in AI research---that scaling general methods outperforms hand-crafted domain knowledge~\cite{sutton2019bitter}.
We offer three responses.
First, even if VLMs eventually learn to interpret any visual encoding, understanding \emph{how current models} process visualizations enables better human-AI collaboration today and reveals properties of machine cognition worth studying independent of practical applications.
Second, the bitter lesson is a heuristic, not a universal law---in practice, the most successful AI systems often combine scaling with domain structure, and the history of science is replete with cases where domain-specific understanding complemented general-purpose methods.
Third, foundation model architectures have proven surprisingly stable: the transformer has dominated since 2017, and VLM vision encoders (CLIP, SigLIP) persist across model generations, suggesting design insights may have longer shelf lives than the rapid-obsolescence argument assumes.

\textbf{Will machines converge on human-like vision?}
We acknowledge evidence that VLMs learn some human-like perceptual structure.
Sanders et~al.~\cite{sanders2025vlm} find that VLMs approximate human similarity geometry for natural images, and VLM-derived psychological spaces can match or exceed the predictive utility of human-derived spaces for some tasks.
Doerig et~al.~\cite{doerig2025highlevel} show that high-level visual representations in the human brain are aligned with LLM embeddings of scene captions---suggesting the brain may encode visual scenes into a semantic format compatible with how LLMs encode language, though notably this alignment is text-mediated rather than a direct visual-to-visual correspondence.
However, this convergence evidence applies primarily to feature processing of natural images.
The high-level constraints that shape visualization design---preattentive feature channels, working memory limits, sequential attention deployment, and the specific perceptual rankings established by Cleveland and McGill---may not emerge from scale alone.
Current benchmark evidence suggests they have not: models show systematic failures on tasks humans find trivial, sensitivity to factors humans ignore, and divergent design preferences~\cite{wang2024dracogpt}.
Whether future architectures will converge on human-like visualization processing remains an open empirical question---one that machine-oriented visualization research could help answer.

\subsection{Implications for Visualization Theory}
This analysis carries three theoretical implications for how the field understands itself.

First, the field's design knowledge should be explicitly recognized as \emph{audience-specific}: existing principles are human perceptual science, valuable and valid but not universal.
This is not a limitation---it is a clarification.
Visualization theory has always been a theory of human visual cognition applied to data representation.
Naming this specificity does not diminish the achievement; it makes the scope precise.
Just as user interface design distinguishes between desktop, mobile, and accessibility contexts, visualization theory can distinguish between human-oriented and machine-oriented design.

Second, expanding visualization's audience requires building new knowledge, not merely applying existing knowledge.
Making the implicit assumption of a human perceiver explicit---and systematically relaxing it---opens a research direction that is theoretical in nature and empirical in method.
The accessibility research community offers a useful precedent here: designing for screen reader users required developing new representation principles, not simply adapting sighted-user designs~\cite{zong2022rich}.
The machine audience case is analogous in structure, though the perceptual differences are arguably larger.

Third, the question of what counts as ``visualization'' may need revisiting.
If machine-oriented visual representations prove valuable but look nothing like conventional charts---if the optimal design for machine cognition is a spatial arrangement that no human would recognize as a visualization---does it still belong within the field's scope?
We suggest that it does, because the core intellectual contribution of visualization research is understanding how visual representation encodes and communicates information.
That contribution is not limited to a single audience.
The field's methods---controlled experiments varying visual design, task-based evaluation, systematic study of encoding effectiveness---are precisely what this new design problem requires.
Expanding the audience does not change what the field does; it extends where the field's expertise applies.

\section{Conclusion}
\label{sec:conclusion}
Visualization's design knowledge is grounded in six decades of human perceptual science.
This knowledge base is rigorous, empirically validated, and enormously valuable---but it was developed for a specific audience.
Benchmark evidence increasingly indicates that it does not straightforwardly transfer to machine audiences: VLMs process charts through different mechanisms, exhibit different patterns of success and failure, and respond to different visual properties than humans do.

The prevailing response---bypassing machine vision with text-based alternatives---is pragmatically sensible but forecloses a research question worth asking: could visual representations designed for machine cognition be valuable?
We do not claim to know the answer.
The evidence is suggestive but early.
What we argue is that the question deserves systematic investigation, and that the field's experimental infrastructure is well-suited to pursue it.

The research program we call for is modest in its immediate claims and ambitious in its scope.
We do not prescribe specific design principles---machine cognition is evolving too rapidly for that, and the optimal forms of machine-oriented visualization remain an open empirical question.
We call instead for the empirical investigation that would be needed to develop such principles: the beginnings of a machine Bertin, a machine Cleveland and McGill.

The visualization field has spent sixty years learning how to design for human minds.
As machines become consumers of visual information, the question of how to design for machine cognition becomes both scientifically interesting and practically relevant.
We are in the early stages of understanding what machine-oriented visualization might mean.
This paper argues that finding out is worth the effort.

\section*{Acknowledgments}
This research is funded by the ANID FONDECYT 11250039 Project. The author is also supported by Project 202311010033-VRIDT-UCN. During the preparation of this work, the authors used Claude to refine sections and support literature review activities. Additionally, Writefull integrated in Overleaf was used to improve writing quality and readability. After using these tools/services, the author reviewed and edited the content as needed and takes full responsibility for the content of the article.

\bibliographystyle{IEEEtran}
\bibliography{references.bib}

\begin{IEEEbiography}[{\includegraphics[width=1in,height=1.25in,clip,keepaspectratio]{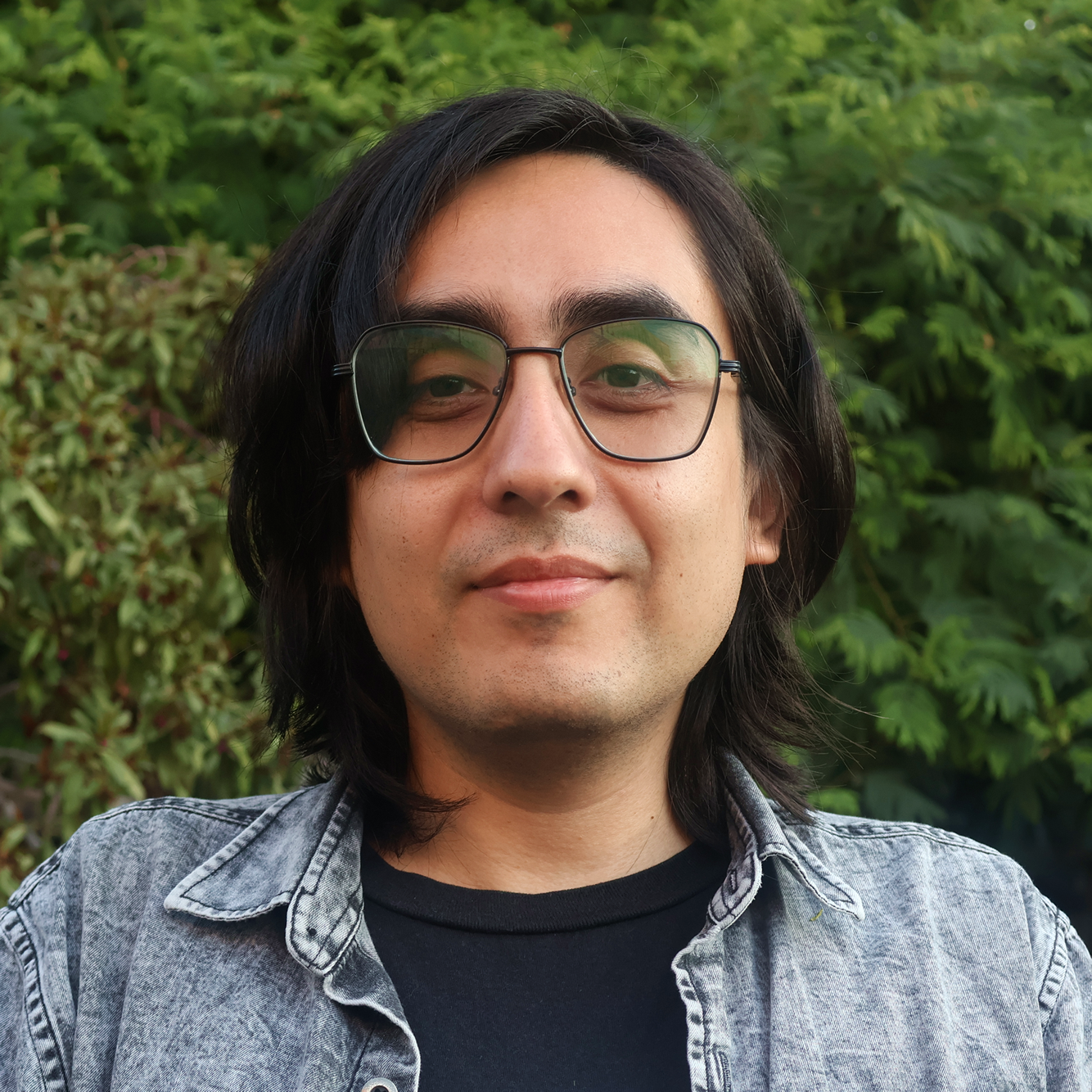}}]{Brian Keith} 
received the B.Sc. degree in engineering and the professional title in computing and informatics civil engineering from Universidad Católica del Norte (UCN), Antofagasta, Chile (2016), the B.Sc. degree in mathematics and the M.Sc. degree in informatics engineering at UCN (2017), and the Ph.D. degree in computer science and applications from Virginia Tech, Blacksburg, VA, USA (2023). He is currently an Associate Professor with the Department of Systems and Computing Engineering and Secretary of Research and Technological Development for the Faculty of Engineering and Geological Sciences at UCN. He is also the Director of the Artificial Intelligence Innovation Center for the Antofagasta Region (CIARA). He is the author of more than 60 research articles. His research interests include visual analytics, artificial intelligence, text analytics, computational narratives, and applied data analytics in geochemistry. Dr. Keith is an Associate Editor of \emph{Intelligent Data Analysis}. He was a recipient of the Fulbright Faculty Development Scholarship (2019-2021), the Becas Chile Doctoral Studies Scholarship (2019-2023).\end{IEEEbiography}

\vfill

\end{document}